# Interpreting Aqueous Two-Phase Extraction of Single-Walled Carbon Nanotubes with Highly Versatile Nonionic Polymers

Błażej Podleśny[a,*], Łukasz Czapura[a], Oussama Er-Riyahi[b], Karolina Z. Milowska[b,c], Dawid Janas[a,*]

[a] Department of Organic Chemistry, Bioorganic Chemistry and Biotechnology, Silesian University of Technology, B. Krzywoustego 4, 44–100 Gliwice, Poland

[b] CIC nanoGUNE, Donostia-San Sebastián 20018, Spain

[c] Ikerbasque, Basque Foundation for Science, Bilbao 48013, Spain

[*] Corresponding authors: Blazej.Podlesny@polsl.pl (B.P), Dawid.Janas@polsl.pl (D.J.)

**Abstract**

The development of efficient separation methods is essential for the production of fine chemicals and materials. Among them, the aqueous two-phase extraction (ATPE) allows for the isolation of single-walled carbon nanotubes (SWCNTs) of specific structures and other substances. However, this easy-to-use method, in which an analyte is partitioned between two phases, still demands a better understanding of its mechanism to make its application more effective. Herein, we demonstrate how various biphasic systems can be formed according to the nature of the phase-forming components. Moreover, by employing polyethylene-block-poly(ethylene glycol) (PEPEG), previously unrecognized in this context, we reveal the versatility of nonionic polymers for ATPE, which can successfully act as phase-forming compounds, partitioning modulators, and dispersing agents. Interestingly, as proven by experiments and modelling, PEPEG exhibited chirality-sensitive preference toward SWCNTs, which can significantly facilitate the purification of SWCNTs using various approaches. Capitalizing on this finding, we report how the extraction environment may be tailored to promote the isolation of (8,3) SWCNTs and other chirality-enriched SWCNT fractions. The relationships noted, based on the examination of a model material (SWCNTs), provide substantial insight into the elusive mechanism of the ATPE purification approach, widely employed across a range of analytes, from cell organelles to nanostructures.

# 1. Introduction

Single-walled carbon nanotubes (SWCNTs) have garnered significant attention over the past three decades due to their unique electrical[1] and optical[2] properties. Although all SWCNTs are made of hexagonal lattices of carbon atoms, they do not form just one strictly defined structure. The cylindrical shape allows for different arrangements of carbon atoms in space, which results in the existence of various species of SWCNT (discriminated using specific (n,m) indices of chirality)[3]. Since the properties of the material are highly dependent on the SWCNT structure[4], control thereof is essential. Unfortunately, SWCNT synthesis leads to polychiral blends containing multiple chiralities[5], so an arsenal of SWCNT sorting techniques has been developed to address this issue. Among these, conjugated polymer extraction[6–8], liquid chromatography[9–11], density gradient ultracentrifugation[12,13], and the aqueous two-phase extraction (ATPE)[14–16] are the most popular methods due to their simplicity and selectivity.

The latter tactic, ATPE, is especially appealing as it relies on the convenient liquid-liquid extraction of SWCNTs between two immiscible aqueous phases[17]. However, as both media are water-based, particular phase-forming compounds need to be introduced at appropriate concentrations to produce such an extraction system. Aqueous two-phase systems were first observed and described by the Dutch researcher Martinus Beijerinck as early as 1896, when he noticed the formation of two phases after mixing agar and gelatin solutions[18]. Sixty years later, the application potential of this scientific curiosity was revealed by Per-Åke Albertsson, who used a dextran (DEX)-poly(ethylene glycol) (PEG) system to purify biological compounds and cell particles. Therefore, Albertsson is considered the creator of the ATPE method[19].

The following years were associated with the development of the ATPE method. New types of two-phase systems have been developed, expanding the choice from the polymer-polymer system to polymer-salt[20], salt-alcohol[21], ionic liquid-salt[22], or ionic liquid-polymer[23]. Attention was paid not only to the measurable effects related to the purification of compounds but also to the laws governing two-phase systems. It was recognized that in order to create a two-phase system, it is crucial for the phase-forming compounds to differ in hydrophilicity (or hydrophobicity) and to reach an appropriate concentration level[17]. The conditions separating the existence of a single- and two-phase system are represented by a binodal curve. In short, the binodal curve indicates the minimum concentration of both components necessary to obtain phase separation (**Figure 1b**). The area above the binodal curve represents the biphasic region, while the area below represents the single-phase region. There are three primary methods for establishing the binodal curve: node determination, turbidimetric titration, and cloud point titration[18]. Depending on the phase-forming agents (and their molar mass), the biphasic region is obtained at different concentrations. Several essential relationships are particularly well-known in polymer-polymer systems. The most important is related to the length of the polymer chains. With increasing chain length, the biphasic region is obtained at lower concentrations than in the case of shorter chain analogs[18]. Furthermore, with increasing temperature or pH, higher concentrations of components are required to get the biphasic state[18].

ATPE is one of the most suitable methods for extracting and purifying species that can be denatured in non-aqueous conditions. The use of ATPE has enabled the purification of cell organelles[24], proteins[25], nucleic acids[26], viruses[27], antibiotics[28], dyes[29], low-molecular-weight organic compounds[30], metals[31], metallic nanoparticles[32], and nanostructures such as SWCNTs[15] or boron nitride nanotubes[33]. Despite these achievements, the separation mechanism of the ATPE method remains incompletely understood. Regardless of the selected system, the separation trend can be expressed, similarly to classical extraction, by means of the partition coefficient K (**Equation 1**), defined as the ratio of the analyte concentration in the top phase ($c_T$) to the analyte concentration in the bottom phase ($c_B$):

$$K = \frac{c_T}{c_B} \quad (1)$$

The subject of discussion remains the factors that influence the partition coefficient. According to the most popular theory of the ATPE mechanism, created by Albertson himself, the partition depends on five main factors[17,18,34]: (1) Hydrophobicity (the hydrophobic properties of both the purified substance and the phase-forming compounds), (2) Specific affinity (the affinity of the purified substance to one of the phase-forming compounds), (3) Electrochemical aspects (the electrochemical potential or electric charge of the purified substance and phase-forming compounds), (4) Size (the size or surface area of the purified substance), and (5) Conformation (the structural arrangement of the purified substance). Taking the above into account, it is assumed that the partition coefficient K can be expressed by (**Equation 2**):

$$lnK = lnK_{hydr} + lnK_{aff} + lnK_{el} + lnK_{size} + lnK_{conf} + lnK_0 \quad (2)$$

Where the symbols *hydr*, *aff*, *el*, *size*, and *conf* denote hydrophobicity, specific affinity, electrochemical, size, and conformation coefficient, respectively, while the $K_o$ factor is related to environmental effects, such as the mass of the phase-forming compound or the degree of solvation of the purified substance. In the case of polymer-polymer and polymer-salt systems, it is assumed that the key factor is hydrophobicity.

The problems associated with understanding the principles governing ATPE partitioning are even more pronounced in the case of SWCNTs, which are not inherently water-dispersible. To suspend them in an aqueous solution, surface-active compounds (surfactants) are necessary. Anionic surfactants, especially bile salts such as sodium cholate (SC) and sodium deoxycholate (SDC), are most commonly used for this purpose due to their exceptional ability to solubilize SWCNTs in water[5]. After introducing the SWCNT dispersion into a two-phase system, the SWCNTs often occupy just one of the phases, which in the case of SC or SDC is typically the bottom phase[15]. To separate SWCNTs into fractions containing SWCNTs of specific structures, the system is tuned by changing the surfactant type/concentration[35] or introducing partitioning modulators[36], the latter of which do not have surface-active properties *per se*. While in the DEX-PEG system SC and SDC tend to push the SWCNTs into the hydrophilic bottom phase enriched in DEX[15], sodium dodecyl sulfate (SDS)[37], sodium dodecylbenzenesulfonate (SDBS)[38], and nonionic surfactants, such as Triton X100[15] and Brij35[39], have an opposite effect, so they can be used to extract SWCNTs to the top phase gradually. Most of the research on SWCNT separation via ATPE exploits the DEX-PEG system[14]. Other combinations, such as isomaltodextrins-PEG[40], polyacrylamide-PEG[41], DEX-polyvinylpyrrolidone[41], and DEX-Pluronic L35[38], are less commonly employed. Therefore, compared to the purification of other aforementioned materials or compounds by ATPE, processing SWCNTs by this method is more complex due to the indispensable presence of surfactants on their surface, which is necessary to make them soluble in water. The selection of appropriate conditions has already enabled the isolation of semiconducting and metallic SWCNT fractions[42], as well as the separation of SWCNTs by diameter[43], chirality[44], and the handedness[45]. Still, the separation mechanism and the role of each compound present in the ATPE system remain unclear.

In this work, we tackle this problem by employing polyethylene-block-poly(ethylene glycol) (PEPEG) copolymer as a model compound to interpret the ATPE mechanics. PEPEG proved to be a highly versatile ATPE component, which, depending on the concentration, acted as a phase-forming agent, a SWCNT partition modulator, and a SWCNT dispersant. Having thoroughly studied the impact of its introduction on SWCNT partitioning, we developed a simple, few-step ATPE method for the isolation of (8,3) SWCNT species. The devised method was, for the first time, very sensitive to the chiral angle of the processed SWCNTs, opening new perspectives for SWCNT separation. Moreover, the obtained results provide considerable insight into extractive processes, which are at the heart of the ATPE method, capable of sorting many compounds and materials, underscoring the interdisciplinary impact of the presented findings.

## 2. Experimental.

All the information about the materials and methods is provided in the SI.

## 3. Results and discussion

### 3.1. Construction of new ATPE systems to study their capacity for two-phase formation

When searching for new two-phase systems for SWCNT separation, we formulated several prerequisites. Knowing that the initial SWCNT dispersion will be prepared in bile salt solutions (SC or SDC), which solubilize them to a great extent[15], the new systems should be of the polymer-polymer type, which are compatible with such an analyte. Choosing the polymer-salt, salt-salt, ionic liquid (IL)-polymer, or salt-IL system would make the separation of SWCNTs challenging if not impossible, as the presence of ions strongly disturbs surfactant-SWCNT interactions, rendering the material unseparable due to its agglomeration[46].

The second requirement was that the new system should behave at least as well as the popular DEX–PEG combination used for the purification of a broad spectrum of compounds and materials[17]. Among nonionic polymers, it is challenging to find an alternative to DEX that can create two-phase systems with PEG and provide a suitable environment for the processing of SWCNTs. The suitability of DEX is related to the presence of 1,6-linkages of glucose in the main chain, resulting in helical geometry of DEX, which seems to be essential during the separation of SWCNTs. According to published works, such a spatial arrangement promotes the differentiation of SWCNTs in an aqueous environment[40]. Hence, since the employment of DEX was considered crucial, we sought new systems of the DEX-polymer type.

The last requirement was related to the concentration level at which the new system exhibits two-phase characteristics. Naturally, for economic and ecological reasons, it is difficult to accept systems wherein 50% of the sample are phase-forming polymers. Additionally, high concentrations of these components would likely increase the viscosity of such biphasic systems considerably, making handling of the material demanding. Therefore, our goal was to identify systems that display the possibility of two-phase formation in the concentration range of 2-15%.

Our initial studies focused on the influence of polymer molecular weights on the shape of the binodal curves that separate one- and two-phase conditions, which we examined using the cloud point titration method with a selection of phase-forming compounds (**Figure 1a**)[47]. **Figure 1b** shows an example binodal curve of a two-phase aqueous system. The X- and Y-axes represent the weight fractions (wt% concentrations) of Polymer 1 (denser, creating the bottom phase) and Polymer 2 (less dense, creating the top phase), respectively. The points on the graph form a binodal curve, indicating the concentrations required to form a two-phase system. For conditions defined above the curve, the system exists as two-phase; below, it is single-phase.

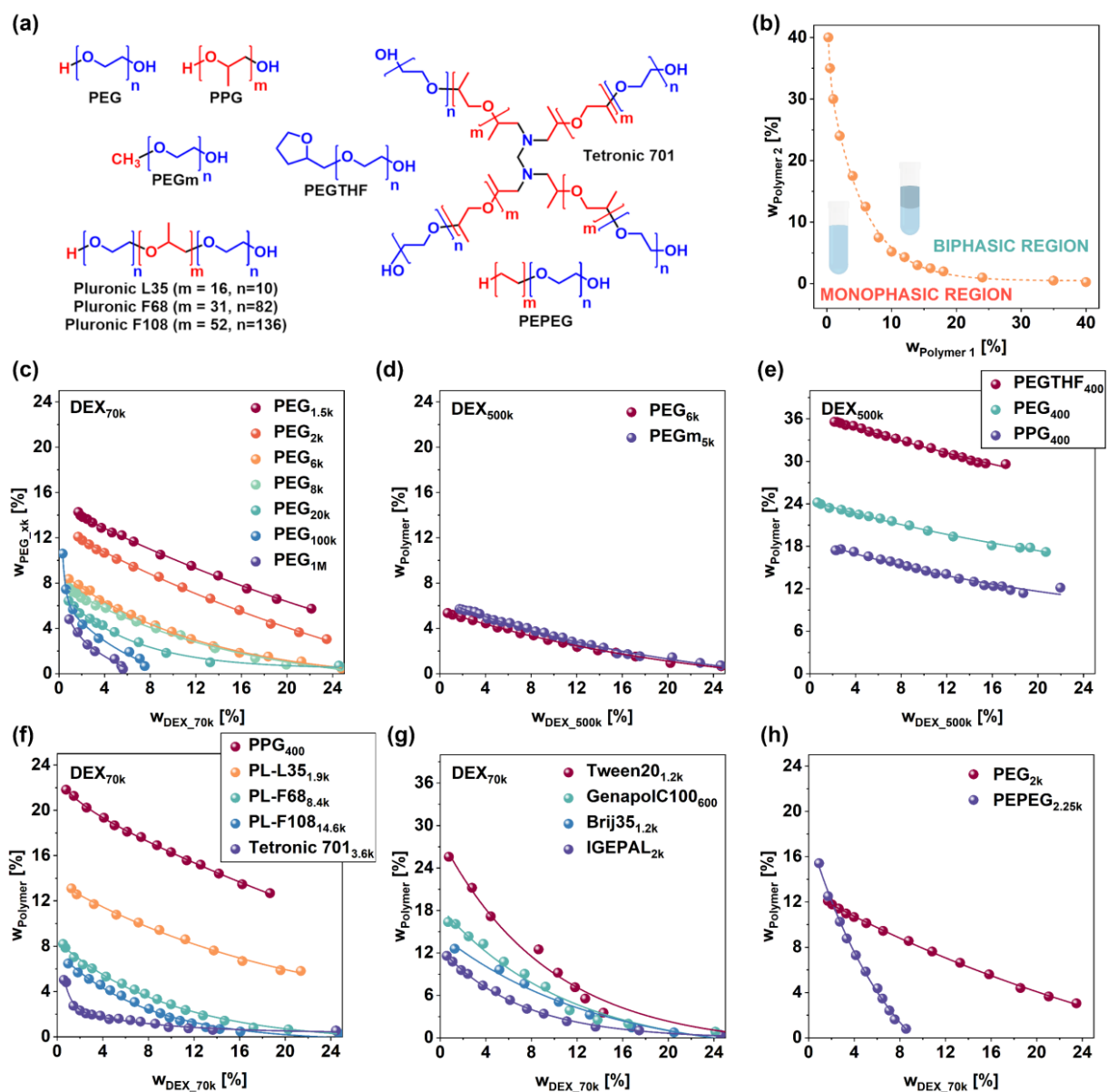

**Figure 1** (a) Structures of the phase-forming compounds used (predominantly hydrophilic and hydrophobic parts are highlighted in blue and red, respectively), (b) example binodal curve depicting the separation of the mono- and bi-phasic conditions, (c) binodal curves of systems containing $DEX_{70k}$ and PEG of various molecular weights, (d) binodal curves of systems containing $DEX_{500k}$ and $PEG_{6k}$ or $PEGm_{5k}$ of comparable molecular weights, (e) binodal curves of systems containing $DEX_{500k}$ and $PEGTHF_{400}$, $PEG_{400}$, or $PPG_{400}$ of comparable molecular weights (please note the difference in range on the Y axis), (f,g) binodal curves of systems containing $DEX_{70k}$ and a selection of polymers (please note the difference in range on the Y axis in panel (g)), and (h) binodal curves of systems containing $DEX_{70k}$ and PEG or PEPEG of comparable molecular weights.

Treating the $DEX_{70k}$-$PEG_{6k}$ system as a reference, we determined its binodal curve. Then, we replaced $PEG_{6K}$ with PEG of various molecular weights. With the elongation of PEG chains giving rise to the increase in PEG's molecular weight, the $DEX_{70k}$-$PEG_{xk}$ binodal curve gradually approached the lower concentration regime (**Figure 1**). Consequently, smaller amounts of large molecular weight polymers were required to create two-phase systems, which conformed to the literature data for $DEX_{10k}$-$PEG_{xk}$ and $DEX_{500k}$-$PEG_{xk}$ systems[48]. For example, to create a system containing 4% $DEX_{70k}$, the concentration of $PEG_{2k}$ must reach 11%. In contrast, the application of $PEG_{20k}$ instead reduces the necessary PEG concentration to about 4%. Hence, the use of polymers with larger molecular weight minimizes the material consumption, making the process more eco-friendly. Other DEX-PEG systems are analyzed in more detail in **Figure S1**.

However, the above-described merit comes at a cost, which is the increased difficulty of routine work involving these biphasic systems due to the increased viscosity of phase-forming solutions of high molecular weights. As a result, analytes separated in such systems require effort, such as vigorous mixing by vortexing, to overcome the diffusion limitations. For this reason, we excluded $PEG_{100k}$ and $PEG_{1M}$ from further use, even though minimal amounts thereof are required to create two-phase systems. Such systems may be operational already at the total weight of both phase-forming components below 2%, which we demonstrated by combining $PEG_{1M}$ and $DEX_{500k}$ (**Figure S1d**).

On the other hand, low-molecular-weight $PEG_{400}$, which is liquid at room temperature, may also be employed for the formation of biphasic systems with DEX. While it can be introduced directly (without water) into the sample to make a two-phase system with DEX, very high concentrations of this compound (exceeding 20%) were required to reach such conditions (**Figure S1e**). This phenomenon may be due to the relatively high hydrophilicity of $PEG_{400}$, which necessitates high concentrations of this component, as the growth of the PEG chain is highly correlated with changes in its hydrophobicity. The longer the PEG chain, the more hydrophobic the polymer is, which is directly related to the number of -OH and -H end groups. Even when the same amount of two PEGs of different molecular weights is used, the shorter polymer has a relatively higher number of end groups in solution, which affects how PEG interacts with DEX and the surrounding liquid environment[49].

To prove that the type of the end group is an essential factor determining the propensity to form biphasic systems, we compared the difference in behavior between two biphasic systems composed of $DEX_{500k}$ and $PEG_{6k}$ or $PEGm_{5k}$ (**Figure 1d**). The former is a methyl ether of PEG, so PEG whose end group is $CH_3$, not H. The binodal curves displayed nearly equal lineshapes, although the PEGm employed had a lower molecular weight (5 kDa) compared to PEG (6 kDa), so, based solely on molecular weight effect, its binodal curve should be positioned higher. However, since the binodal curves of both described systems matched, the results suggested that the more hydrophobic character of PEG may have counteracted the molecular weight effect.

To support the hypothesis that the degree of hydrophilicity affects the shape of the binodal curve, we extended the scope of analyzed biphasic systems and included two additional phase-forming components. In addition to $PEG_{400}$, more hydrophobic (poly(propylene glycol), $PPG_{400}$) and hydrophilic (tetrahydrofurfuryl alcohol poly(ethylene glycol) ether, $PEGTHF_{400}$) were employed in conjunction with $DEX_{500k}$ (**Figure 1e**). The most hydrophobic PPG required the lowest concentrations to form a two-phase system (**Figure S1e**), which was consistent with previously interpreted dissimilarities between $PEG_{6k}$-$DEX_{500k}$ and $PEGm_{5k}$-$DEX_{500k}$ systems (**Figure 1d**). The results of these experiments validated this reasoning and were a good hint for searching for a new phase-forming agent that could replace PEG.

Many commercially available compounds can be regarded as PEG derivatives of potential use for the development of the ATPE technique. Among such compounds are PEG copolymers with polypropylene glycol (Pluronic (PL) variant), polyoxyethylenes (Triton, Brij, Genapol, and IGEPAL variant), or ethoxylated sorbitans (Tween variant). Additionally, most of them have surface-active properties [50], which may improve the resolution of biphasic systems for the separation of materials and compounds. Some of these compounds have already been used to create two-phase systems with DEX[51], but the literature lacks data on the conditions under which these compounds can generate biphasic systems. This knowledge gap requires attention, since, as we recently demonstrated, biphasic systems based on Dextran and Pluronic may be used for the separation of SWCNTs[38]. Several important conclusions were drawn based on evaluating biphasic systems made of $DEX_{70k}$ and $PPG_{400}$, PL-L35$_{1.9k}$, PL-F68$_{8.4k}$, PL-F108$_{14.6k}$, Tetronic701$_{3.5k}$, Tween20$_{1.2k}$, GenapolC100$_{600}$, Brij35$_{1.2k}$, or IGEPAL$_{2k}$ (**Figure 1fg**).

$DEX_{70k}$ formed stable biphasic systems with all of the examined compounds, which had various molecular weights. Among the obtained results, the binodal curves of two Pluronic compounds (PL-F68$_{8.4k}$ and PL-F108$_{14.6k}$) with $DEX_{70k}$ stand out. Despite the significant difference in molecular weight of these nonionic polymers, their binodal curves showed a good match. These block copolymers, composed of hydrophobic poly(propylene oxide) (PPO) core flanked by hydrophilic poly(ethylene oxide) (PEO) chains, had the same relative ratio of the number of hydrophobic to hydrophilic units, suggesting that PL-F68$_{8.4k}$ and PL-F108$_{14.6k}$ had a similar degree of hydrophilicity. This factor again seems to be crucial for establishing biphasic conditions with other phase-forming compounds.

Equally interesting was the binodal curve of the DEX$_{70k}$-Tetronic701$_{3.5k}$ system, showing that a small concentration of this nonionic compound is sufficient to form a biphasic system with dextran. Structural analysis of this compound explains why this is the case and supports our earlier claims. In Tetronic701, the number of hydrophobic PPO parts per chain is 14, while the number of hydrophilic PEO moieties amounts to just 2[52]. Consequently, the nature of this compound is predominantly hydrophobic, and, as we earlier observed, the binodal curves of hydrophobic compounds tend to occupy the small-concentration regime. Besides the scientific value of these findings, systems such as DEX and Tetronic701 have not been reported and may potentially have good application potential, due to the presence of nitrogen atoms in Tetronic701, which can interact with numerous analytes to be separated.

**3.2. Analysis of the role of ATPE components in the extractive purification of materials**

Encouraged by these results, we also successfully constructed a biphasic system from dextran and polyethylene-block-poly(ethylene glycol) (PEPEG$_{2.25k}$, PEG:PE ratio 80:20), which is distinctly more hydrophobic than unmodified PEG. The presence of a hydrophobic PE part significantly reduced the concentrations needed to achieve a biphasic system, compared to PEG$_{2k}$ (**Figure 1h**). Additionally, PEPEG at a high concentration (16%) was easy to work with in the laboratory (no foaming, low viscosity), so we employed it for ATPE of SWCNTs, which in this work, served as a model material to elucidate the mechanism of the process.

SWCNT separation was performed in the DEX$_{70k}$-PEPEG$_{2.25k}$ system, where DEX$_{70k}$ constituted 8.33% (m/V) and, previously unused in this context, PEPEG$_{2.25k}$ 6.67% (m/V) of the total sample volume. Such parameters allowed for the split into two phases, with a bottom-to-top phase volume ratio of 2:1 (total sample volume was chosen to be 4.8 mL). Within this system, we partitioned small-diameter raw SWCNTs suspended in 2% SC aqueous solution (**Table S3**). The studies began with a sample composed of only phase-forming polymers and the SWCNT dispersion (no surfactant added to facilitate SWCNT partitioning). To our surprise, practically all of the SWCNTs occupied the top phase (**Figure 2a**). Typically, SC-suspended SWCNTs occupy the DEX-rich phase in the commonly exploited DEX-PEG systems[15].

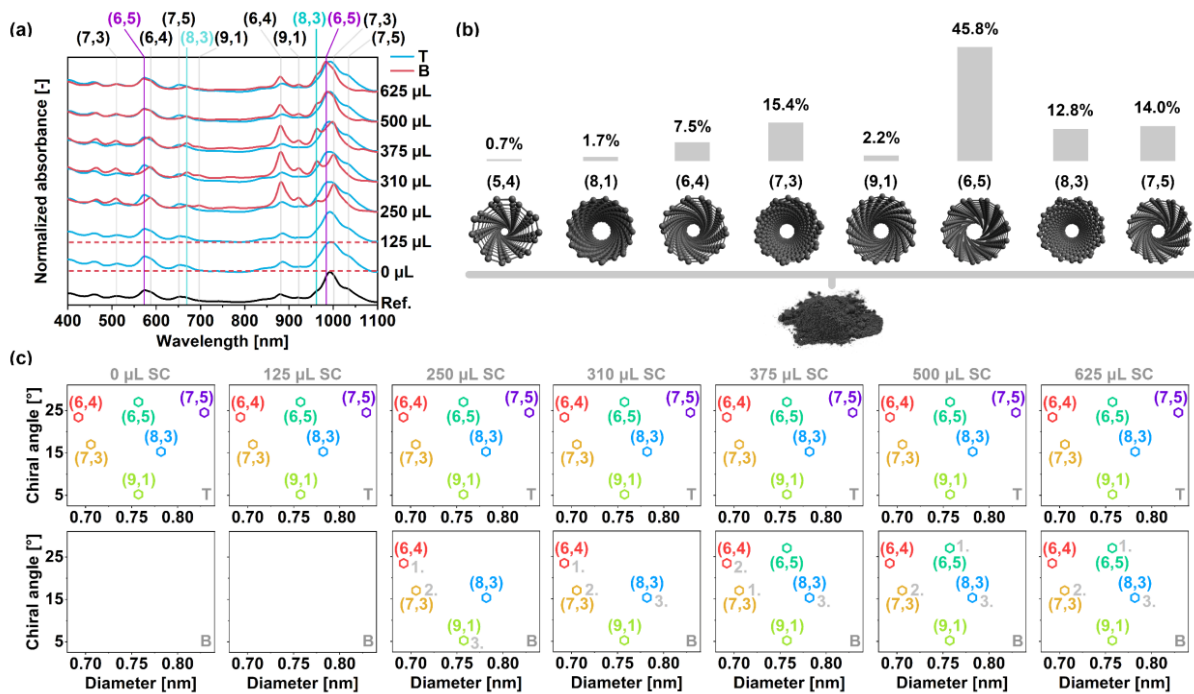

**Figure 2** (a) Optical absorption spectra of top and bottom phases obtained in a DEX$_{70k}$-PEPEG$_{2.25k}$ systems using a dispersion of small-diameter SWCNTs in 2% SC aqueous solution and various indicated volumes of 40% SC aqueous solution (SWCNT-free fractions are indicated with dashed lines), (b) composition of small-diameter SWCNTs[53] sorted in this experiment (SWCNTs are arranged in ascending diameter order from left to right), (c) the presence of indicated SWCNT species detected in the top and bottom phases of the aforementioned extraction system as a function of volume of added SC (the three most abundant SWCNT chiralities are indicated with numbers for the bottom phases). The absorbance is normalized to the intensity of the E$_{11}$ optical transition of the major SWCNT chirality to facilitate data comparison.

Interestingly, in preliminary research, we noted that the addition of a relatively concentrated SC (10%) did not cause migration of SWCNTs to the bottom phase at all even for 500 μL of SC, which was the maximum possible amount of SC to be added to preserve the constant sample volume of 4.8 mL. Only when even more concentrated SC (40%) was used did the SWCNTs start to emerge in the bottom phase (**Figure 2a**). With an increase in SC concentration in the system, small-diameter SWCNTs initially migrated to the bottom phase, and the increase in the SC concentration caused the migration of gradually larger SWCNT species to the bottom phase. After the introduction of 250 μL 40% SC additive, peaks of (6,4) (perhaps the smallest of semiconducting SWCNTs in the material[15]), (7,3), and (9,1) SWCNTs appeared in the bottom phase. Increasing the SC addition to 310 μL resulted in the appearance of clearly visible peaks of (8,3) SWCNT in the bottom phase. Further increasing the SC amount in the system (375-500 μL) promoted the transition of (6,5) SWCNTs to the bottom phase. The maximum amount of SC tested by us (625 μL), breaking the constant sample volume regime (**Table S3**), caused the migration of a larger amount of SWCNTs to the bottom phase. However, most of the SWCNTs were still in the top phase, and the composition of this phase barely changed even in the presence of a high content of SC.

This series of experiments provided us with valuable information, shedding more light on the elusive mechanism of the ATPE process. Firstly, the partitioning of SWCNTs in the newly designed $DEX_{70k}$-$PEPEG_{2.25k}$ system is possible, so the derivative of PEG maintains its capacity to form a biphasic system with DEX. Secondly, PEPEG has a distinctly different nature than PEG typically employed for top-phase formation. While the introduction of moderate amounts of SC to the DEX-PEG system readily moves all the SWCNTs to the bottom phase[15,46], a drastically higher concentration of SC in the $DEX_{70k}$-$PEPEG_{2.25k}$ system still does not allow for complete migration of SWCNTs to the bottom phase. The alkyl chain introduced to PEG, which is present in PEPEG, makes the top phase even more hydrophobic than bare PEG already is[15]. Due to the hydrophobic nature of SWCNTs, they have a stronger interaction with the PEPEG environment, which, due to the presence of the PE units, is also notably hydrophobic. Consequently, it is more challenging to promote the downward migration of SWCNTs. The third observation results from the qualitative composition of the isolated SWCNT fractions. With the addition of bile salt surfactants such as SC to a DEX-PEG system, SWCNTs always diffuse to the bottom phase, starting with those of the smallest diameters[35,46]. Surprisingly, we noted that not all SWCNTs followed this relation in the DEX-PEPEG system. Despite the fact that (6,5) SWCNTs (d = 0.757 nm) account for almost half of the raw material **(Figure 2b)**, they were detected in the bottom phase after (8,3) SWCNTs (d = 0.782 nm), which emerged there first, even though their diameter is larger. However, (9,1) SWCNTs, which are of the same diameter as (6,5) SWCNTs but with a different chiral angle, migrated to the bottom phase as expected, i.e., following the diameter order. Hence, the results suggest that PEPEG introduces a chiral angle preference to the ATPE system, which, when fine-tuned, may be exploited to separate hard-to-resolve SWCNTs of similar chiral angles. The distinctly increased affinity to (6,5) SWCNTs, which keeps it in the top phase, as will be shown in the further parts of this work, is a vital feature of PEPEG as it helps to avoid its extraction to the bottom phase. In a DEX-PEG system, this effect is highly unfavorable as it complicates SWCNT purification and frequently contaminates other monochiral SWCNT fractions, given that (6,5) SWCNTs account for nearly half of the composition of the raw SWCNTs. The application of sodium deoxycholate (SDC) instead of SC produced analogous results, but, due to its stronger capacity to promote SWCNT migration[15,46], much lower concentrations of SDC were required for the SWCNT differentiation to occur (**Table S4, Figure S2**).

To better understand how PEPEG influences the ATPE, we returned to a typical DEX-PEG separation system and evaluated whether this compound may positively influence SWCNT partitioning when used as an additive, not as a phase-forming agent. The $DEX_{70k}$-$PEG_{6k}$ system, containing equal concentrations (5.88% m/V) of both components, a dispersion of small-diameter SWCNTs in 2% SC aqueous solution, and 10% SC aqueous solution to move them initially to the bottom phase, and 2% PEPEG (100 µL, 300 µL, 900 µL, 1800 µL) to gradually extract the SWCNTs to the top phase were combined (**Table S5**). A small amount of PEPEG (100 µL) already promoted the partitioning of the SWCNTs between two phases (**Figure 3a**), with the (6,5) SWCNTs unfortunately dominating in both phases. Still, the increased amount of PEPEG (300 µL) drastically reduced the content of (6,5) in the bottom phase, which was moved to the top. Concomitantly, the bottom phase was enriched with (6,4), (7,3), (9,1), and (8,3). The emerging presence of the latter could again be a surprise, considering that these SWCNTs have a larger diameter than (6,5) SWCNT, which migrated to the top (**Figure 3b**). Based solely on diameter differences between these species, the larger (8,3) SWCNTs should have diffused to the top phase readily than (6,5) SWCNTs. With further increase of PEPEG volume to 900 µL and 1800 µL, (8,3) and (9,1) SWCNTs were detected in the top phase, respectively. In parallel, the bottom phase was enriched with (7,3) SWCNTs. The results of this experiment provide more concrete evidence that PEPEG does exhibit improved affinity toward (6,5) SWCNTs, accompanied by a simultaneous reduction in preference for (8,3) SWCNTs, unlocking the possibility of chiral angle-driven separation of SWCNTs with ATPE.

To confirm the unexpected affinity of $PEPEG_{2.25k}$ toward (6,5) SWCNT, we performed an analogous experiment related to the $PEPEG_{2.25k}$ additive series using different starting material (HiPco SWCNTs). It has a different composition and a larger diameter distribution, with (6,5) SWCNT not being the dominant fraction[44]. The smallest volume of the $PEPEG_{2.25k}$ additive examined (100 µL) showed that the major species in the bottom phase was (6,5) SWCNT (**Figure 3c**) with other chralities also present, such as (8,3), (9,1) SWCNT, (7,3), (6,4), (7,5) and (10,2). Increasing the $PEPEG_{2.25k}$ amount in the system to 300 µL resulted in the migration of the larger species ((7,5) SWCNT d=0.829 nm and (10,2) SWCNT d=0.884 nm) from the bottom phase. (6,5) SWCNTs remained as the primary chirality in the bottom phase, while the relative amount of the (8,3) SWCNTs increased, which again should not be the case for diameter-controlled SWCNT partitioning[54].

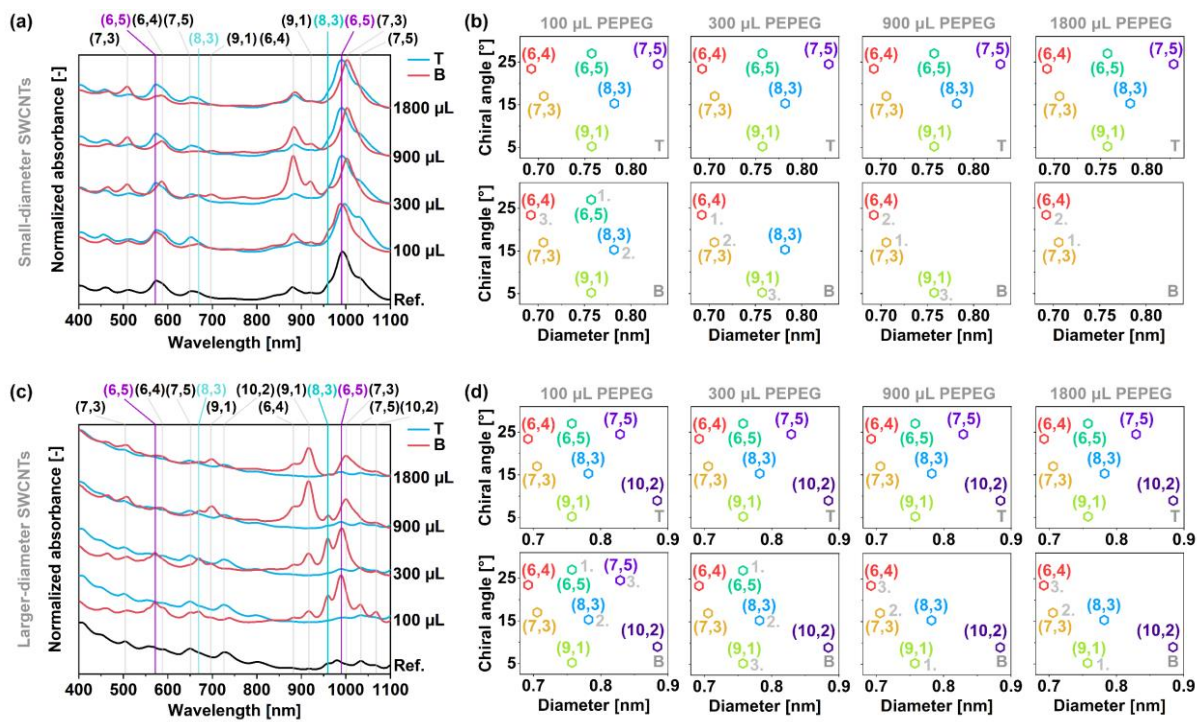

**Figure 3** (a) Optical absorption spectra of top and bottom phases obtained in a DEX$_{70k}$-PEG6$_k$ systems using a dispersion of small-diameter SWCNTs in 2% SC and various indicated volumes of 2% PEPEG$_{2.25k}$, (b) the presence of indicated SWCNT species detected in the top and bottom phases of the aforementioned extraction system as a function of volume of added PEPEG (the three most abundant SWCNT chiralities are indicated for the bottom phases). (c) optical absorption spectra of top and bottom phases obtained in a DEX$_{70k}$-PEG6$_k$ systems using a dispersion of larger-diameter SWCNTs in 2% SC and various indicated volumes of 2% PEPEG, (d) the presence of indicated SWCNT species detected in the top and bottom phases of the aforementioned extraction system as a function of volume of added PEPEG (the three most abundant SWCNT chiralities are indicated for the bottom phases). The absorbance is normalized to the intensity of the E$_{11}$ optical transition of the major SWCNT chirality to facilitate data comparison.

The addition of 900 μL PEPEG$_{2.25k}$ caused complete migration of the (6,5) SWCNTs from the bottom to the top phase, and the main fraction in the former became (9,1) SWCNT. This is significant as (6,5) SWCNTs have precisely the same diameter as (9,1) SWCNTs, yet the latter did not diffuse to the top at the indicated amount of the PEPEG additive. Besides, the level of (9,1) enrichment is unprecedented, especially considering that this SWCNT type is found in negligible concentration in the raw material of ca. 2% (**Figure 2b**). Furthermore, the spectrum also shows peaks corresponding to (6,4), (7,3), and (8,3) chiralities. Further increase in the PEPEG$_{2.25k}$ solution volume to 1800 μL caused the (8,3) SWCNT peak to disappear from the bottom phase spectrum. These results strengthened our earlier stated claim that PEPEG$_{2.25k}$ exhibits improved affinity for the near-armchair (6,5) SWCNT. At the same time, the near-zig-zag (9,1) SWCNTs seemed virtually unextractable by the PEPEG-based surfactant mixture.

Likewise, across the tested processing conditions, the bottom phases always contain low-chiral-angle uncommon (10,2) SWCNTs, which somehow cannot migrate to the top phase even though they have the largest diameter in the investigated spectral range (**Figure 3d**).

The results suggest that PEPEG$_{2.25k}$ may have surface-active properties, which promote the extraction of certain SWCNTs. Considering its structure, the hydrophobic polyethylene parts undoubtedly have an affinity for the hydrophobic SWCNT surface via van der Waals interactions. At the same time, the poly(ethylene glycol) segments likely prefer interaction with the aqueous environment, as PEG is a compound well-known for being hydrophilic[55]. Consequently, the PEPEG$_{2.25k}$ may act as a surfactant, because surfactants commonly used to facilitate the extraction of SWCNTs with ATPE from one phase to the other behave analogously[15,39]. In addition, the structural similarity of PEPEG$_{2.25k}$ to the top-phase forming component PEG may explain why the addition of this compound causes the upward migration of SWCNTs.

Furthermore, it is essential to determine the role that phase-forming components themselves may play in extraction. To examine whether the chemical identity and concentration of phase-forming components can also affect the course of SWCNT separation, a biphasic system composed of DEX$_{70k}$ and PL-F68$_{8.4k}$ was formed, and the concentration of PEG$_{70k}$ was increased from 0 to ca. 4%. Interestingly, as the PEG concentration increased, the amount of SWCNTs in the top phase also increased slightly but gradually (**Figure S4**). The content of small-diameter SWCNTs decreased stepwise from the bottom phase, while larger SWCNTs migrated to the top phase. However, this effect was much less pronounced compared to when PEPEG was used as an additive in the previous section. With a high likelihood, this is related to the discovered surface-active properties of PEPEG, increasing the affinity of this compound to SWCNTs. On the contrary, PEG is not equipped with a hydrophobic PE tail, which can interact with SWCNTs. Yet, even in its absence, PEG caused a discernible influence on SWCNT extraction.

Furthermore, considering the potential high affinity of PEPEG$_{2.25k}$ to SWCNTs, we investigated whether it can also serve as a dispersing agent for SWCNTs. Unfortunately, the SWCNT dispersion prepared entirely in 2% PEPEG$_{2.25k}$ had a poor quality and precipitated spontaneously even before centrifugation. Therefore, we created bisurfactant solutions composed of PEPEG$_{2.25k}$ and a specific amount of SC to improve the stability of the SWCNT dispersion. Regardless of the amount of PEPEG$_{2.25k}$ introduced, the spectra of all dispersions of raw SWCNTs were analogous, which indicated that the same types of SWCNTs were solubilized for every dispersant combination examined. Compared to pure SC@SWCNT dispersion, the peaks displayed by SC-PEPEG$_{2.25k}$@SWCNT suspensions were not as sharp, meaning that the SWCNTs were less well individualized[56] (**Figure 4a**). The observed peak shape is reminiscent of that of SWCNT material dispersed in nonionic surfactants **(Figure S3)**, supporting our hypothesis that PEPEG$_{2.25k}$ behaves like one. The fact that the optical transitions in the spectra were redshifted also indicated partial agglomeration of the material[57] Nonetheless, the obtained results supported our reasoning that PEPEG$_{2.25k}$ is an active component of the biphasic system, as it readily deposits on the surface of SWCNTs in such a dispersion.

To validate the utility of the bisurfactant SWCNT dispersion, we used it as feed for SWCNT separation using the ATPE method. A suspension prepared in the SC:PEPEG$_{2.25k}$ (3:1) mixture was introduced into the DEX$_{70k}$-PEG$_{6k}$ system in the presence of water and pure SC solution to promote the division of SWCNTs between the phases (**Table S8**). The resulting top phase was black, while the lower exhibited an orange-gray color. The intense color of the former suggested that most of the material was located in the top phase, which was confirmed by spectroscopy. The recorded optical spectrum of the top phase resembled the unsharp spectrum of the starting material (**Figure 4b**). On the contrary, the spectrum of the bottom phase was of very good quality, and the peaks were well resolved. It mainly comprised (6,4), (7,3), (8,3), and (9,1) SWCNTs. The bottom sample generated this way was further separated by mixing it in a volume ratio of 1:2 with the prepared corresponding complementary top phase, consisting of only PEG, DEX, PEPEG, and SC. This procedure allowed for the enrichment of the (8,3) SWCNT species (with (6,4) and (9,1) SWCNT contamination) in the top phase and isolation of the (7,3)- and (9,1)-rich SWCNT fractions (with (8,3) and (9,1) SWCNT contamination) in the bottom phase. Finally, the 3$^{rd}$ extraction step was performed to purify (8,3) SWCNTs further. For this purpose, the top obtained from the second extraction step was combined again with a complementary top phase free of SWCNTs, the details of which are provided in the Supplementary Information. As a result, most of the (8,3) SWCNTs migrated to the top phase, giving a fraction highly enriched with this chirality, as evidenced by the enclosed optical absorption spectrum (**Figure 4c**). The recorded photoluminescence excitation-emission map corroborated high isolation selectivity. A tiny amount of (6,4) SWCNTs was not detected in the PL map as the emission profile of this SWCNT type extends beyond the detection range of our instrument. Nonetheless, no other SWCNT species were detected in the emission window of 950-1650 nm, which validates a high value of the harvested SWCNTs for photonics.

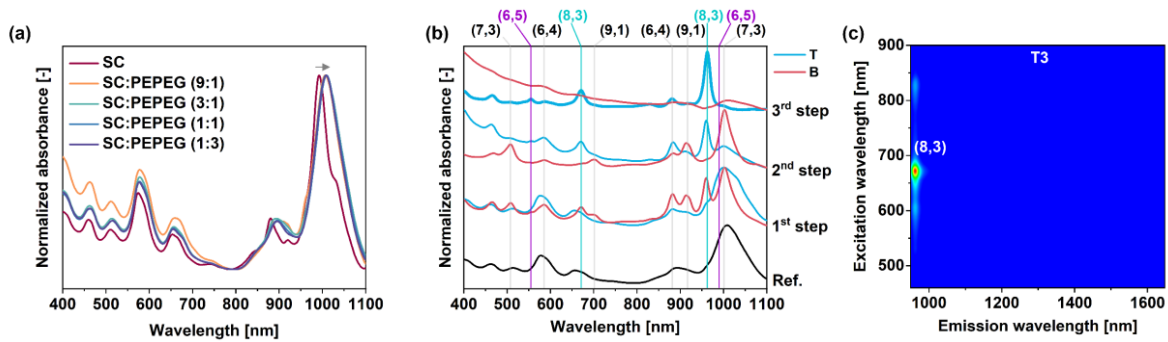

**Figure 4** (a) Optical absorption spectra of dispersions of small-diameter SWCNTs in SC and SC-PEPEG$_{2.25k}$ mixtures at the indicated weight ratios, (b) optical absorption spectra of top and bottom phases obtained in a three step separation using a DEX$_{70k}$-PEG6$_k$ system and SC-PEPEG$_{2.25k}$ @SWCNT (3:1) dispersion, (c) excitation-emission photoluminescence map of the top phase from the third extraction step.

### 3.3. Modeling of the chirality-sensitive interaction of PEPEG with SWCNTs

To elucidate the differing interactions between PEPEG and (6,5) versus (8,3) SWCNTs, we performed density functional theory (DFT), molecular dynamics (MD), and time-stamped force-bias Monte Carlo (MC) simulations. We began by analyzing the frontier orbitals of two amphiphilic polymers, PE$_4$PEG$_9$ and PE$_8$PEG$_{18}$, in the context of their interactions with (6,5) and (8,3) SWCNTs (**Figure 5a**), using spin-polarized DFT with a general gradient approximation (GGA) exchange–correlation functional[58] and dispersive correction[59]. The calculated HOMO and LUMO energy levels, aligned with the projected density of states (DOS) of (6,5) and (8,3) SWCNTs, revealed that the shorter PEPEG polymer is likely to form type-I heterojunctions with both SWCNTs. In contrast, the longer polymer appears more consistent with type-II band alignment. Furthermore, the spatial distribution of the HOMO and LUMO orbitals in the short polymer suggests localization mainly within the PEG segment, particularly at the chain termini (LUMO), whereas for PE$_8$PEG$_{18}$, the HOMO localizes on the PE segment and the LUMO remains on PEG. Since the band structures of (6,5) and (8,3) SWCNTs are relatively similar—both with comparable band gaps and aligned valence and conduction band edges—electronic level alignment alone does not explain differences in polymer–SWCNT interaction. Therefore, we performed MD/MC simulations of both SWCNTs with three PEPEG chains homogeneously distributed around each SWCNT in aqueous solution.

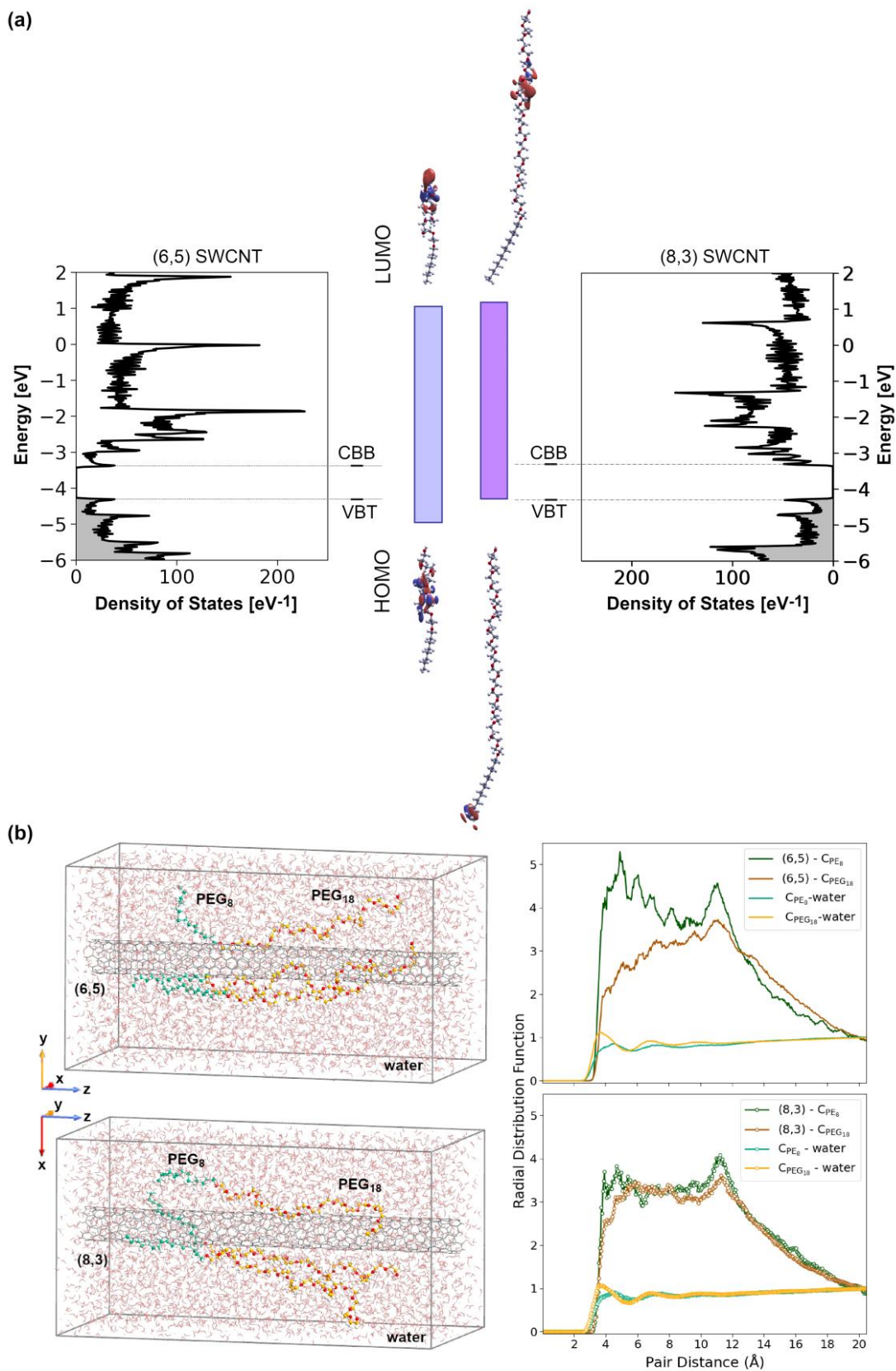

**Figure 5** (a) Energy alignment diagram showing the projected density of states (PDOS) for (6,5) and (8,3) SWCNTs and the HOMO/LUMO energy levels of PEPEG polymers of varying lengths in vacuum, calculated at the SP-DFT/PBE+D2/PZD level of theory. All energy values are presented on an absolute energy scale. The valence band top (VBT) and conduction band bottom (CBB) of each SWCNT are indicated. HOMO and LUMO energy levels of the fully relaxed short and long PEPEG polymers are depicted as vertical bars. The spatial distributions of the HOMO and LUMO orbitals are visualized at an isovalue of 0.02 Å$^{-3/2}$, with red and blue lobes representing the negative and positive phases of the wavefunction, respectively. Carbon, oxygen, and hydrogen atoms are colored grey, red, and white, respectively. (b) Results of MD/MC simulations of SWCNT–polymer interactions in aqueous solution. Snapshots show the final configurations for simulation boxes containing PE$_8$PEG$_{18}$ polymers interacting with either two units of (6,5) SWCNT (top left) or two units of (8,3) SWCNT (bottom left), both immersed in water. SWCNTs, solvent, and polymer molecules are represented using stick, line, and ball-and-stick models, respectively. Solvent molecules are rendered with higher transparency for clarity. Carbon atoms in the PE and PEG segments are shown in teal and yellow, respectively. RDF plots (right) display the pair distribution functions between SWCNT carbon–PE carbon, SWCNT carbon–PEG carbon, PE carbon–water, and PEG carbon–water atom pairs.

These simulations (**Figure 5b**) showed that the hydrophobic PE block preferentially adsorbs near the lateral surface of the SWCNT, whereas the hydrophilic PEG block is more exposed to water. This spatial separation along the polymer chain is consistent with its amphiphilic character and has direct implications for interaction with the SWCNT surface. This behavior is confirmed by radial distribution function (RDF) analysis—quantifying the probability of finding specific atom pairs at a given distance—shown in the right panel of **Figure 5b**. The position of the first peak in the PEG–water RDF occurs at shorter distances than the first peak in the PE–water RDF, indicating stronger PEG–water affinity. Moreover, the polymer conformation differs significantly in vacuum (**Figure 5a**) compared to aqueous solution (**Figure 5b**). The PEG segment becomes more extended when solvated, whereas it appears condensed in vacuum. This result confirms the relative hydrophilicity of this fragment. Concomitantly, in the SWCNT–polymer RDFs, the higher magnitude and shorter distance of the PE–SWCNT peak relative to the PEG–SWCNT peak further support the preferential proximity of the PE block to the SWCNT surface.

When comparing the two systems, distinct differences emerge between the (6,5) and (8,3) cases. For (6,5) SWCNTs (top panel of **Figure 5b**), PE segments show significantly stronger interaction with the SWCNT surface than PEG segments. In contrast, the distinction is less pronounced in the (8,3) SWCNT system (bottom panel). Shorter polymers exhibit similar trends (**Figure S5**), with even greater differences in peak magnitudes for PE–SWCNT and PEG–SWCNT RDFs, resembling the (6,5) system more closely. Regardless of polymer length, the first RDF peak corresponding to PE–SWCNT interaction is consistently shifted to slightly longer distances in the (8,3) SWCNT system, suggesting weaker or more spatially diffuse interactions compared to the (6,5) case. This shift highlights that PE chains tend to position themselves further from the (8,3) SWCNT than from the (6,5) SWCNT. The preferential alignment of the PE segment along the (6,5) SWCNT observed in the simulations may promote more stable polymer–SWCNT interactions, facilitating the migration of (6,5) SWCNTs to the top phase. PEG and PEPEG are structurally similar, which may increase the affinity between these two compounds. Hence, the stronger binding of PEPEG to (6,5) SWCNTs enhances the likelihood of transferring SWCNTs of chirality to the top phase. Conversely, the larger (8,3) SWCNTs should diffuse to the top phase more readily because of their larger diameter, but the less stable interaction of this chirality with PEPEG makes it unfavorable.

## 4. Conclusions

ATPE is a highly effective method for purifying various substances, including polychiral SWCNTs. However, the underlying mechanism is not fully understood, hindering its more widespread utilization. The elaboration of the behavior of a broad spectrum of newly constructed ATPE systems enabled us to deduce how the nature of the phase-forming compounds affects their capacity for generating biphasic systems. We noted that the difference in hydrophilicity between the two phases substantially affects the propensity for creating a biphasic system and its subsequent operation. Consequently, the obtained insight paves the way toward more conscious design of biphasic systems for purification of chemical compounds and materials.

Moreover, we observed that phase-forming compounds also influence the extraction process. The more surface-active the chemical is, the stronger its influence on the extraction course. For instance, the addition of a simple hydrophilic PEG makes a negligible impact on the SWCNT distribution in the biphasic system. In contrast, the more amphiphilic PEPEG (containing the SWCNT-prefering PE moiety) alters SWCNT extraction to a much higher degree. Based on the knowledge obtained through experimentation and modeling, we designed an effective method for the separation of (8,3) SWCNTs. Interestingly, the inclusion of PEPEG in the biphasic system made the ATPE separation sensitive to the chiral angle of the processed SWCNTs. This discovery introduces a new dimension to SWCNT separation by ATPE and other techniques, such as gel chromatography or ultracentrifugation, opening the route to the precise harvesting of SWCNTs with desired characteristics from complex polychiral SWCNT mixtures.

**Acknowledgments**

The authors would like to thank the Polish National Science Centre for the financial support of the research (under the OPUS program, Grant agreement 2019/33/B/ST5/00631). The contribution of Ms. Aneta Suska is appreciated. K.Z.M. and O. R. gratefully acknowledge the Interdisciplinary Centre for Mathematical and Computational Modelling at the University of Warsaw, Poland (Grant No. G47-5), DIPC Supercomputing Center, Spain, and the Theory of Condensed Matter Group at the Cavendish Laboratory, University of Cambridge, UK, for providing computer facilities and technical support. K.Z.M. also would like to thank the European Commission (Marie Skłodowska-Curie Cofund Programme; grant no. H2020-MSCA-COFUND-2020-101034228-WOLFRAM2) and Agencia Estatal de Investigación, Ministerio de Ciencia e Innovación, Spain (Proyectos de Generación de Conocimiento 2022 program, PID2022-139776NB-C65) for funding this research.

**Conflict of interest**

The authors declare no competing interests.

Supplementary Information file

# Interpreting Aqueous Two-Phase Extraction of Single-Walled Carbon Nanotubes with Highly Versatile Nonionic Polymers

Błażej Podleśny[a,*], Łukasz Czapura[a], Oussama Er-Riyahi[b],
Karolina Z. Milowska[b,c], Dawid Janas[a,*]

[a] *Department of Organic Chemistry, Bioorganic Chemistry and Biotechnology, Silesian University of Technology, B. Krzywoustego 4, 44–100 Gliwice, Poland*

[b] *CIC nanoGUNE, Donostia-San Sebastián 20018, Spain*

[c] *Ikerbasque, Basque Foundation for Science, Bilbao 48013, Spain*

[*] Corresponding authors: Blazej.Podlesny@polsl.pl (B.P), Dawid.Janas@polsl.pl (D.J.)

**Table of Contents**

# 1. Experimental

## 1.1. Materials

Signis SG65i (CoMoCAT65, Sigma-Aldrich, USA, batch: MKCK1004) and HiPco (NanoIntegris, Canada; batch HP30-006, purified) SWCNTs were used for separation experiments.

Sodium cholate (SC, PanRecAppliChem, Germany), sodium deoxycholate (SDC, Sigma-Aldrich, USA), sodium chenodeoxycholate (SCHDC, Abcr, Germany), sodium taurodeoxycholate (STDC, Abcr, Germany), Pluronic F-127 (Pluronic F127, Sigma-Aldrich, USA), Polyethylene-block-poly(ethylene glycol) (PEPEG, Sigma-Aldrich, USA), Tween20 (VWR, USA), GenapolC100 (Sigma-Aldrich, USA), BrijS100 (Sigma-Aldrich, USA), IGEPAL CO-890 (IGEPALCO890, Sigma-Aldrich, USA) were used to prepare SWCNT dispersions.

The compounds used to prepare aqueous two-phase systems are listed in **Table S1**. **Table S1** also contains data provided by the manufacturers regarding the average molar mass of the polymers. $M_W$, $M_R$, and $M_N$ represent weight average, relative, and number average molar mass, respectively. We also present "working concentration", i.e., the one at which the compound was still well-soluble in water, and the solution obtained allowed for comfortable work (no gelation, precipitation out, etc.).

For all experiments and characterization, double-distilled water obtained from the Elix Millipore system was used.

**Table S1**. Compounds that were used to prepare two-phase systems, their vendors, acronyms, and working concentrations.

| Compound | Molar mas [Da] | Vendor | Acronym | Working concentration (w/w, water) [%] |
|---|---|---|---|---|
| Dextran | 40 000 (-) | PanRecAppliChem (Germany) | $DEX_{40k}$ | 25 |
| Dextran | 70 000 ($M_R$) | PanRecAppliChem (Germany) | $DEX_{70k}$ | 25 |
| Dextran | 250 000 ($M_W$) | Alfa Aesar (Germany) | $DEX_{250k}$ | 25 |
| Dextran | 500 000 ($M_W$) | Alfa Aesar (Germany) | $DEX_{500k}$ | 25 |
| Poly(ethylene glycol) | 400 ($M_N$) | Sigma-Aldrich (USA) | $PEG_{400}$ | 100 |
| Poly(ethylene glycol) | 1 500 ($M_N$) | Alfa Aesar (Germany) | $PEG_{1.5k}$ | 50 |
| Poly(ethylene glycol) | 2 000 ($M_N$) | Alfa Aesar (Germany) | $PEG_{2k}$ | 50 |
| Poly(ethylene glycol) | 6 000 ($M_N$) | Alfa Aesar (Germany) | $PEG_{6k}$ | 50 |
| Poly(ethylene glycol) | 8 000 ($M_N$) | PanRecAppliChem (Germany) | $PEG_{8k}$ | 50 |
| Poly(ethylene glycol) | 20 000 ($M_N$) | Alfa Aesar (Germany) | $PEG_{20k}$ | 40 |
| Poly(ethylene oxide) | 100 000 ($M_W$) | Alfa Aesar (Germany) | $PEG_{100k}$ | 11 |
| Poly(ethylene oxide) | 600 000 - 1 000 0000 ($M_W$) | Thermo Fisher Scientific (USA) | $PEG_{1M}$ | 5 |
| Poly(ethylene glycol) methyl ether | 5 000 ($M_N$) | Sigma-Aldrich (USA) | $PEGm_{5k}$ | 50 |
| Tetrahydrofurfuryl alcohol polyethyleneglycol ether | 400 (-) | Sigma-Aldrich (USA) | $PEGTHF_{400}$ | 100 |
| Polypropylene glycol | 400 (-) | Sigma-Aldrich (USA) | $PPG_{400}$ | 100 |
| Brij35 | 1 199 (-) | Sigma-Aldrich (USA) | $Brij35_{1.2k}$ | 20 |
| Triton X-100 | 602-646 (-) | Thermo Fisher Scientific (USA) | $TX100_{600}$ | 25 |
| Pluronic P-123 | 5 800 ($M_N$) | Sigma-Aldrich (USA) | $PluronicP123_{5.8k}$ | 25 |
| Pluronic F-127 | 12 600 (-) | Sigma-Aldrich (USA) | $PluronicF127_{12.6k}$ | 20 |
| Pluronic F-68 | 8 400 ($M_N$) | Sigma-Aldrich (USA) | $PluronicF68_{8.4k}$ | 25 |
| Pluronic L-35 | 1 900 ($M_N$) | Sigma-Aldrich (USA) | $PluronicL35_{1.9k}$ | 25 |
| Pluronic F-108 | 14 600 ($M_N$) | Sigma-Aldrich (USA) | $PluronicF108_{14.6k}$ | 20 |
| Igepal CO-890 | 1 982 ($M_N$) | Sigma-Aldrich (USA) | $IGEPALCO890_{2k}$ | 20 |
| Tween 20 | 1 128 (-) | VWR (USA) | $T20_{1.2k}$ | 20 |
| Tetronic 701 | 3 600 ($M_N$) | Sigma-Aldrich (USA) | $Tetronic701_{3.6k}$ | 20 |
| Polyethylene-block-poly(ethylene glycol) | 2 250 ($M_N$) | Sigma-Aldrich (USA) | $PEPEG_{2.25k}$ | 16 |
| GenapolC100 | 627 (-) | Sigma-Aldrich (USA | $GenapolC100_{600}$ | 20 |

## 1.2. Optical characterization

UV–VIS–NIR spectra were measured with the Hitachi U2910 spectrophotometer. Excitation-emission photoluminescence (PL) maps were measured with a ClaIR spectrophotometer (Photon etc., Canada) across the specified wavelength ranges: excitation (460–900 nm) and emission (948–1600 nm).

## 2. Modelling

### 2.1. DFT calculations

The spin-polarized Density Functional Theory (DFT)[1,2] calculations for (6,5) and (8,3) SWCNTs, PE$_4$PEG$_9$, and PE$_8$PEG$_{18}$ polymers were carried out using a generalized gradient approximation (GGA) employing a PBE[3] exchange and correlation functional with D2 Grimme's correction[4] for dispersive interactions, and double-ζ plus polarization numerical basis (DZP) sets, as implemented in SIESTA numerical package[5,6]. The use of a commonly employed exchange-correlation functional instead of a hybrid functional was motivated by the fact that hybrid functionals have been shown to poorly describe the size-dependent evolution of HOMO-LUMO gaps in polymers, due to their tendency to over-delocalize molecular orbitals[7,8]. DFT calculations for the finite polymer systems were conducted in 'molecule' mode, employing a sufficiently large simulation box to eliminate spurious interactions between periodic images. The Brillouin zone was sampled exclusively at the Γ point, and the real-space integration grid was defined with a density mesh cut-off of 350 Ry. For the SWCNT systems, the simulation cell dimensions perpendicular to the SWCNT axis were expanded to prevent inter-tube interactions. Brillouin zone sampling along the SWCNT axis was enhanced using a (1 × 1 × 5) Monkhorst-Pack k-point grid[9]. All the structures were relaxed until the maximum force acting on any atom was lower than 0.02 eV/Å and the maximum stress changed by less than 0.001 GPa, while the self-consistent field (SCF) cycle was iterated until the total energy changed by less than $10^{-5}$ eV, and the density matrix elements differed by less than $10^{-4}$ per iteration. The solvent was not included in these calculations.

### 2.2. MD and MC calculations

To investigate the interactions between PEPEG and (6,5) SWCNT, as well as between PEPEG and (8,3) SWCNT in aqueous environments, we performed a series of molecular dynamics (MD) and time-stamped force-bias Monte Carlo (TFMC)[10,11] simulations of infinite SWCNTs interacting with polymers in water (**Figure 5 and Figure S5**). Due to the use of three-dimensional periodic boundary conditions, the simulation boxes contained only two units of either (6,5) or (8,3) SWCNT. Two PEPEG polymers of different lengths—PE$_4$PEG$_9$ and PE$_8$PEG$_{18}$—were initially positioned symmetrically around the SWCNTs, aligned along the SWCNT axis, with approximately equal distances between both the PE and PEG segments and the lateral surface of the SWCNT. The simulation boxes—sized to prevent direct interactions between periodic images of the SWCNT–polymer complexes (4.5 × 4.5 × 8.4 nm³ for systems containing (8,3) SWCNT and 4.5 × 4.5 × 8.1 nm³ for those with (6,5) SWCNT)—were filled with water using Packmol[12]: 4493 molecules for the (8,3)+shorter polymer system, 4480 for (8,3)+longer polymer, and 4491 for systems containing (6,5) SWCNT with both polymers.

MD/MC calculations were performed using a full periodic table bonded valence forcefield - a Universal Force Field (UFF) potential[13], as implemented in QuantumATK[14,15]. Energy contributions to the UFF potential were represented by simple functions based on bond lengths, bond angles, torsion angles, inversion angles, and inter-atomic distances. The electrostatic interactions were calculated using smooth-particle-mesh-Ewald (SPME) solver[16]. The cut-off used for calculating the real-space interactions was set to 7.5 Å, while the relative accuracy of SPME summation was set to 0.0001. Atomic partial charges on each atom were assigned using QEq charge equilibration method[17]. Dispersive interactions were included in the form of Lennard-Jones potential[18–20] with 10 Å cut-off and 2 Å smoothing length.

Four different systems were simulated: (i) an (8,3) SWCNT interacting with three $PE_4PEG_9$ polymers, (ii) an (8,3) SWCNT interacting with three $PE_8PEG_{18}$ polymers, (iii) a (6,5) SWCNT interacting with three $PE_4PEG_9$ polymers, and (iv) a (6,5) SWCNT interacting with three $PE_8PEG_{18}$ polymers, all immersed in water. Each system underwent a short NVT simulation using a Berendsen thermostat[21] (time step: 0.01 fs; total time: 1 ps; thermostat coupling constant: 1000 fs), followed by an NPT simulation using the Martyna–Tobias–Klein barostat and thermostat[22] (time step: 0.01 fs; total time: 20 ps; thermostat coupling: 1000 fs; barostat coupling: 5000 fs) at 300 K. Random initial velocities were assigned to all atoms according to the Maxwell–Boltzmann distribution.

Following the production MD simulations, a brief geometry optimization was performed using the LBFGS algorithm for 20 000 steps[23]. The systems were then subjected to TFMC (time-stamped force-bias Monte Carlo) simulations to capture longer-timescale phenomena. TFMC simulations were carried out under both NVT and NPT conditions using the Berendsen thermostat and barostat at 300 K and 1 bar, over a total simulation time of 6.3 ns. The maximum atomic displacement per Monte Carlo step was set to 0.05 Å. The system compressibility was defined as 0.0001 bar$^{-1}$, and the barostat coupling constant was set to 500. Radial distribution functions (RDFs) were computed using data collected during the final 1 ns of the simulations.

# 3. Aqueous two-phase system

## 3.1. Determination of binodal curves

The magnetic stirring bar was placed in a Nessler cylinder, and the weight of the clean and dry apparatus was recorded. Then, a clean stock solution of DEX (25% w/w) was introduced into the cylinder, and the weight of the entire system was recorded. Next, a clean stock solution of the second phase-forming polymer of known concentration (w/w) was added dropwise to the DEX solution and stirred with a magnetic stirrer until the resulting suspension became cloudy for at least 10 seconds. The weight of the entire system was recorded. Then, water was added dropwise to the cloudy suspension while stirring until a clear solution was formed and remained in this state for at least 10 seconds. The weight of the entire system was re-recorded. The procedure leading to a cloudy/transparent solution was repeated several times to collect more measurement points. The data obtained were used to calculate the mass fractions of both polymers necessary to get a two-phase system and to plot the binodal curve. All experiments were conducted at room temperature.

## 3.2. Biphasic systems composed of DEX and PEG

Binodal curves made with phase-forming components of various molecular weights, i.e., PEG of gradually increased molecular weight ($PEG_{1.5k}$, $PEG_{2k}$, $PEG_{6k}$, $PEG_{8k}$, $PEG_{20k}$), were combined with $DEX_{40k}$, $DEX_{250k}$, and $DEX_{500k}$ in **Figure S1a**, **Figure S1b,** and **Figure S1c**, respectively. Replacing $DEX_{70k}$, as described in the main text, with DEX of a lower mass ($DEX_{40k}$) increased the PEG concentrations needed to obtain the biphasic conditions (**Figure S1a**). The situation was reversed when using heavier DEX fractions (250 kDa and 500 kDa, **Figure S1bc**). Interestingly, the difference in behavior between $DEX_{250k}$ and $DEX_{500k}$ was relatively small, and the binodal curves with PEG of the same molecular weight had very similar lineshape. Furthermore, PEG of large (1 MDa) and small (400 Da) molecular weight required low and high concentrations of DEX, respectively, to form biphasic systems (**Figure S1de**). Again, the difference in behavior between $DEX_{250k}$ and $DEX_{500k}$ was relatively small (**Figure S1e**). Finally, the results display that $DEX_{500k}$ may be used to form biphasic systems with various nonionic polymers (**Figure S1f**).

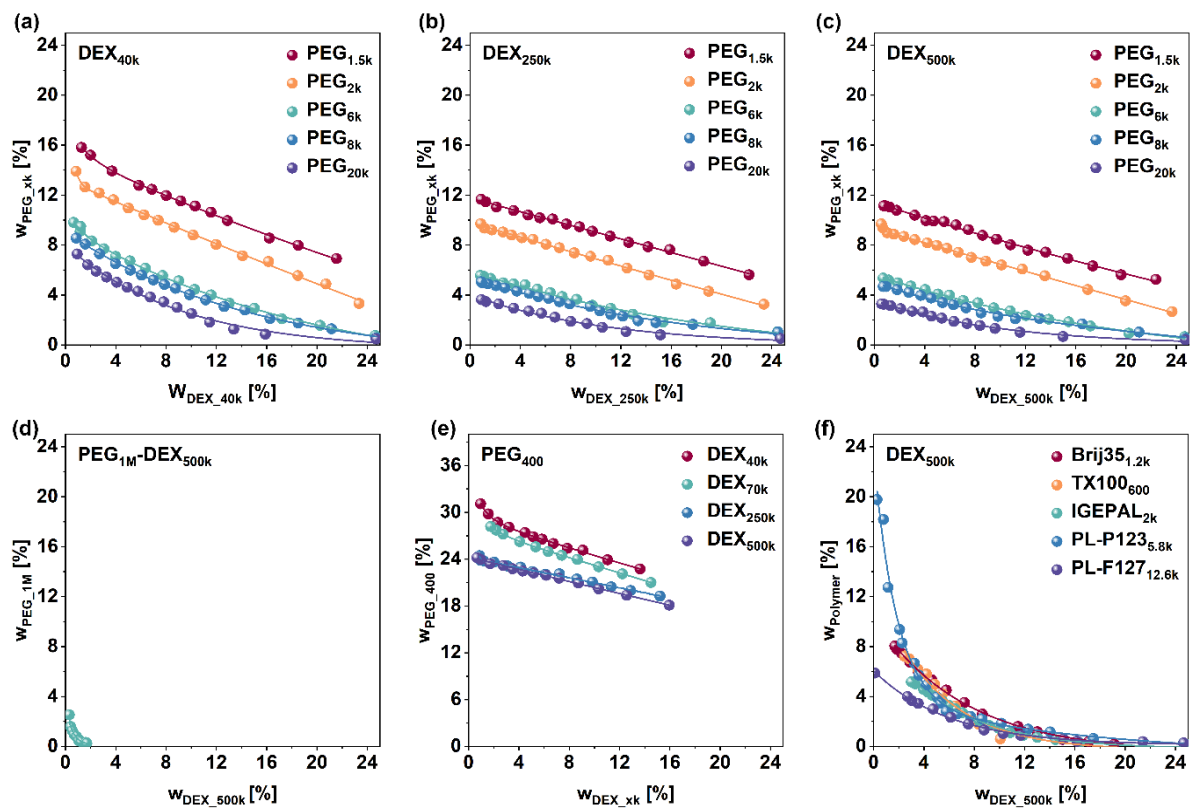

**Figure S1**. Binodal curves of systems composed of (a) $DEX_{40k}$, (b) $DEX_{250k}$, and (c) $DEX_{500k}$ with PEG of various molecular weights. (d) Binodal curve of a system made of heavy molecular weight $PEG_{1M}$ and $DEX_{500k}$. (e) Binodal curves of systems based on $PEG_{400}$ and DEX of various molecular weights (please note the difference in range on the Y axis), (f) binodal curves of systems containing $DEX_{500k}$ and a selection of polymers.

## 4. Separation of SWCNTs

### 4.1. Preparation of SWCNT dispersions

First, a surfactant solution containing 2% (w/w) was prepared. In the case of single-surfactant dispersions, it was made of a 2% solution of a given component. For the SC:PEPEG mixture, the total combined surfactant concentration was 2%, but the concentrations of individual components were different, depending on the SC:PEPEG ratios examined. The concentrations of the compounds in the sample are presented in **Table S2**.

Then, 40 mL of as-prepared surfactant solution (2%) and 40 mg of SWCNTs (CoMoCAT65 or HiPco) were introduced into a 50 mL vial. Then, the mixture, which was kept in an ice bath, was sonicated (Hielscher UP200St, 200 min, 30 W) to solubilize the SWCNTs. Next, the freshly prepared suspension was centrifuged (Eppendorf Centrifuge 5804 R) at 18 °C at the Relative Centrifugal Force (RCF) of 15314 × g for 1.5 h to precipitate bundled SWCNTs. The upper 80% of supernatants were collected and used in the experiments described below.

**Table S2** SC:PEPEG ratios and concentrations of SC and PEPEG in the dispersions prepared in a bisurfactant system.

| SC:PEPEG ratio | Concentration of SC in dispersion [%] | Concentration of PEPG in dispersion [%] | Total surfactant concentration in dispersion [%] |
|---|---|---|---|
| 9:1 | 1.8 | 0.2 | 2 |
| 3:1 | 1.5 | 0.5 | |
| 1:1 | 1 | 1 | |
| 1:3 | 0.5 | 1.5 | |

## 4.2. ATPE protocol

*First step:* Stock solutions of phase-forming polymer (DEX and PEPEG or PEG), SWCNT dispersion, surfactants (SC, SDC, PEPEG), and water (if needed) were transferred to a centrifuge tube. The mixture was homogenized by a Vortex mixer (about 10 s per sample) and centrifuged (Eppendorf Centrifuge 5804 R) for 3 minutes at 18 °C at an RCF of 2025 × g to facilitate two-phase formation. The top and bottom phases were collected by pipetting.

*Second step:* The bottom phase solution obtained after the first step was mixed with a complementary top phase free of SWCNTs in a volume ratio of 1:2. The mixture was then processed analogously as above.

*Third step:* The top phase solution obtained after the second step was mixed with a complementary bottom phase free of SWCNTs in a volume ratio of 2:1. The mixture was then processed analogously as in the previous step.

## 4.3. SWCNT partitioning in the DEX$_{70k}$-PEPEG$_{2.25k}$ system

All experiments using the DEX$_{70k}$-PEPEG$_{2.25k}$ system were one-step separations. Regardless of the SWCNT dispersion used (prepared in 2% SC or SDC), each sample contained 2 mL of DEX$_{70k}$ solution (20%, w/w), 2 mL of PEPEG$_{2.25k}$ solution (16%, w/w), and 300 μL of SWCNT dispersion. The total volume of all samples (excluding the case of 625 μL of 40% SC additive described elsewhere) was 4.8 mL. The sample compositions for the SC and SDC experimental series are provided in **Table S3** and **Table S4**, respectively.

**Table S3** Compositions of SWCNT samples obtained using the $DEX_{70k}$-$PEPEG_{2.25k}$ system, SWCNT dispersion prepared in SC aqueous solution (2%), and SC aqueous solution (40%). The volumes of cholate solution and water introduced are given in the table.

| $DEX_{70k}$ (20%) [μL] | $PEPEG_{2.25k}$ (16%) [μL] | SWCNT dispersion (2% SC) [μL] | SC (40%) [μL] | $H_2O$ [μL] |
|---|---|---|---|---|
| 2 000 | 2 000 | 300 | - | 500 |
| 2 000 | 2 000 | 300 | 125 | 375 |
| 2 000 | 2 000 | 300 | 250 | 250 |
| 2 000 | 2 000 | 300 | 310 | 190 |
| 2 000 | 2 000 | 300 | 375 | 125 |
| 2 000 | 2 000 | 300 | 500 | - |
| 2 000 | 2 000 | 300 | 625 | - |

**Table S4** Compositions of SWCNT samples obtained using the $DEX_{70k}$-$PEPEG_{2.25k}$ system, SWCNT dispersion prepared in SDC aqueous solution (2%), and SDC aqueous solution (5%). The volumes of deoxycholate solution and water introduced are given in the table.

| $DEX_{70k}$ (20%) [μL] | $PEPEG_{2.25k}$ (16%) [μL] | SWCNT dispersion (2% SDC) [μL] | SDC (5%) [μL] | $H_2O$ [μL] |
|---|---|---|---|---|
| 2 000 | 2 000 | 300 | 100 | 400 |
| 2 000 | 2 000 | 300 | 200 | 300 |
| 2 000 | 2 000 | 300 | 300 | 200 |
| 2 000 | 2 000 | 300 | 400 | 100 |
| 2 000 | 2 000 | 300 | 500 | - |

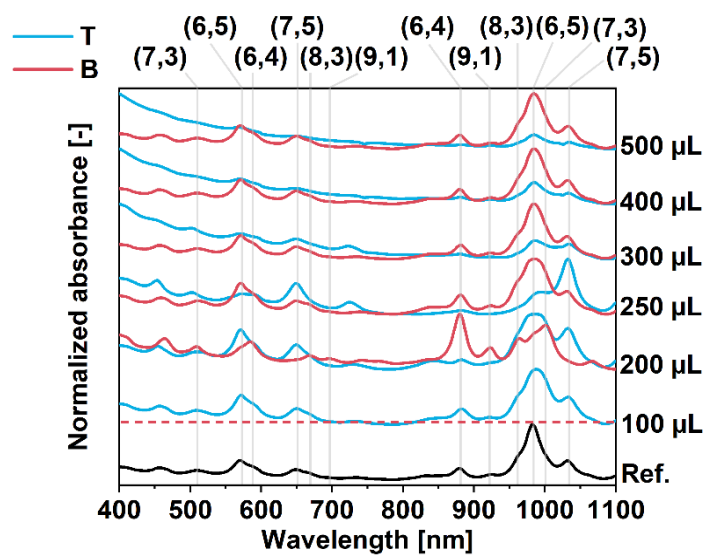

**Figure S2** SWCNT partitioning in the $DEX_{70k}$-$PEPEG_{2.25k}$ system. Optical absorption spectra of the top and bottom phases were obtained after sorting small-diameter raw SWCNTs (SG65i) dispersed in SDC aqueous solution (2%). SWCNT-free fraction is indicated with a dashed line. The absorbance is normalized to the intensity of the $E_{11}$ optical transition of the major SWCNT chirality.

## 4.4. SWCNT partitioning in the DEX$_{70k}$-PEG$_{6k}$ system

All experiments using the DEX$_{70k}$-PEG$_{6k}$ system, SC solution, PEPEG$_{2.25k}$ solution, and SWCNT dispersions (prepared in 2% SC) were one step. Regardless of the starting material (SG65i or HiPco), the composition of the studied ATPE systems was identical. Each sample contained 1350 µL of DEX$_{70k}$ solution (20% w/w), 540 µL of PEG$_{6k}$ solution (50%), 360 µL of SC solution (10%), and 225 µL of SWCNT dispersion. The total volume of all samples was 4590 µL. The sample compositions are provided in **Table S5.**

**Table S5** Compositions of SWCNT samples separated using the DEX$_{70k}$-PEG$_{6k}$ system in one step. The volumes of PEPEG solution and water introduced are given in the table.

| DEX$_{70k}$ (20%) [µL] | PEG$_{6k}$ (50%) [µL] | SC (10%) [µL] | SWCNT dispersion (2% SDC) [µL[ | PEPEG$_{2.25k}$ (2% w/w) [µL] | H$_2$O [µL] |
|---|---|---|---|---|---|
| 1350 | 540 | 360 | 225 | 100 | 2015 |
| 1350 | 540 | 360 | 225 | 300 | 1815 |
| 1350 | 540 | 360 | 225 | 900 | 1215 |
| 1350 | 540 | 360 | 225 | 1800 | 315 |

## 4.5. Evaluation of the quality of SWCNT suspensions made with various dispersants

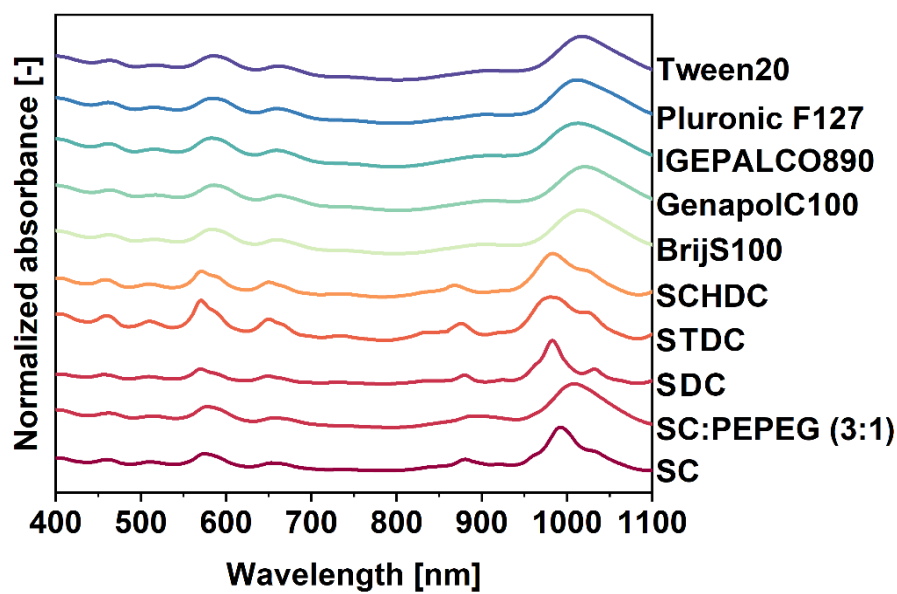

**Figure S3** Optical absorption spectra of (6,5)-enriched raw SWCNTs (SG65i) dispersed in selected surfactant solutions, i.e., sodium cholate (SC), sodium deoxycholate (SDC), sodium taurodeoxycholate (STDC), sodium chenodeoxycholate (SCHDC), a 3:1 mixture of SC and PEPEG, BrijS100, IGEPAL CO-890, Pluronic F-127, and Tween20. The absorbance is normalized to the intensity of the $E_{11}$ optical transition of the major SWCNT chirality to facilitate data comparison.

## 4.6. Evaluation of the role of PEG additive on the separation of SWCNTs in a biphasic system composed of DEX$_{70k}$ and PL-F68$_{8.4k}$

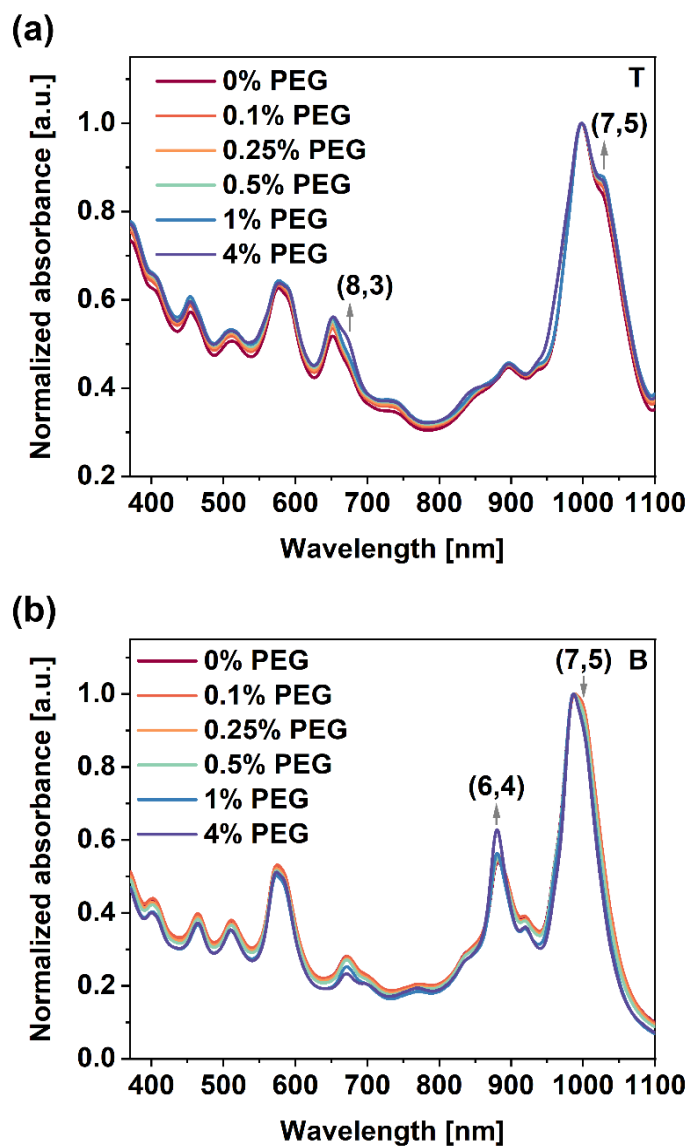

**Figure S4** Optical absorption spectra of (6,5)-enriched raw SWCNTs (SG65i) dispersed in SC (2%) aqueous solution sorted in a biphasic system made of DEX$_{70k}$ and PL-F68$_{8.4k}$ in the presence of various amounts of PEG (listed below in Table S6). The absorbance is normalized to the intensity of the E$_{11}$ optical transition of (6,5) SWCNTs.

The composition of each biphasic system, which enabled acquisition of the above-shown spectra **(Figure S4)**, is given in **Table S6**. The results show that even the addition of PEG causes small but discernible migration of SWCNTs from the bottom to the top phase. While the content of large-diameter (7,5) SWCNTs decreases in the bottom phase, the abundance of smaller species such as (6,4) SWCNTs increases therein.

**Table S6** Compositions of SWCNT samples separated using the DEX$_{70k}$-PL-F68$_{8.4k}$ system in one step with varied concentrations of PEG.

| DEX$_{70k}$ (20%) [μL] | PL-L68$_{8.4k}$ (25%) [μL] | SC (10%) [μL] | TX100$_{600}$ (1%) [μL] | SWCNT dispersion (2% SC) [μL] | PEG$_{6k}$ (50% w/w) [μL] | H$_2$O [μL] |
|---|---|---|---|---|---|---|
| 450 | 360 | 120 | 50 | 75 | 0 | 475 |
| 450 | 360 | 120 | 50 | 75 | 3 | 472 |
| 450 | 360 | 120 | 50 | 75 | 7.5 | 467.5 |
| 450 | 360 | 120 | 50 | 75 | 15 | 460 |
| 450 | 360 | 120 | 50 | 75 | 30 | 445 |
| 450 | 360 | 120 | 50 | 75 | 120 | 355 |

**4.7. Three-step partitioning of SWCNTs in the DEX$_{70k}$-PEG$_{6k}$ system**

The three-step extraction process involved further use of the phase obtained after the first step, followed by the second step. To get a two-phase system in the second and third steps, we used complementary phases prepared analogously (free of SWCNT dispersion but containing DEX$_{70k}$, PEG$_{6k}$, SC, and water). The preparation details are provided in **Table S7**. A higher amount of SC was used in the third step to facilitate the transfer of the SWCNT material to the bottom phase.

**Table S7** Concentrations of stock solutions employed for the three-step ATPE SWCNT partitioning.

| Compound | Concentration of stock solution (w/w; water) [%] |
|---|---|
| DEX$_{70k}$ | 20 |
| PEG$_{6k}$ | 50 |
| SC | 10 |
| PEPEG$_{2.25k}$ | 2 |
| H$_2$O | 100 |

**Table S8** General composition of samples from three-step extraction experiments

| Stock solution | Volume of stock solution [μL] | | | | |
| --- | --- | --- | --- | --- | --- |
| | First step | Second step | | Third step | |
| | | Complementary phase | Extraction system at the second step | Complementary phase | Extraction system at the third step |
| $DEX_{70k}$ | 1350 | 1350 | - | 1350 | - |
| $PEG_{60k}$ | 540 | 540 | - | 540 | - |
| SC | 390 | 405 | - | 2700 | - |
| $PEPEG_{2.25k}$ | - | 300 | - | - | - |
| SWCNT dispersion | 225 | - | - | - | - |
| $H_2O$ | 2085 | 1995 | - | - | - |
| Bottom phase (first step) | - | - | 1500 | - | - |
| Top phase (blank sample, second step) | - | - | 3000 | - | - |
| Top phase (second step) | - | - | - | - | 3000 |
| Bottom phase (blank sample, third step) | - | - | - | - | 1500 |

## 4.8. Effect of PEPEG length on chirality-sensitive interactions with SWCNTs

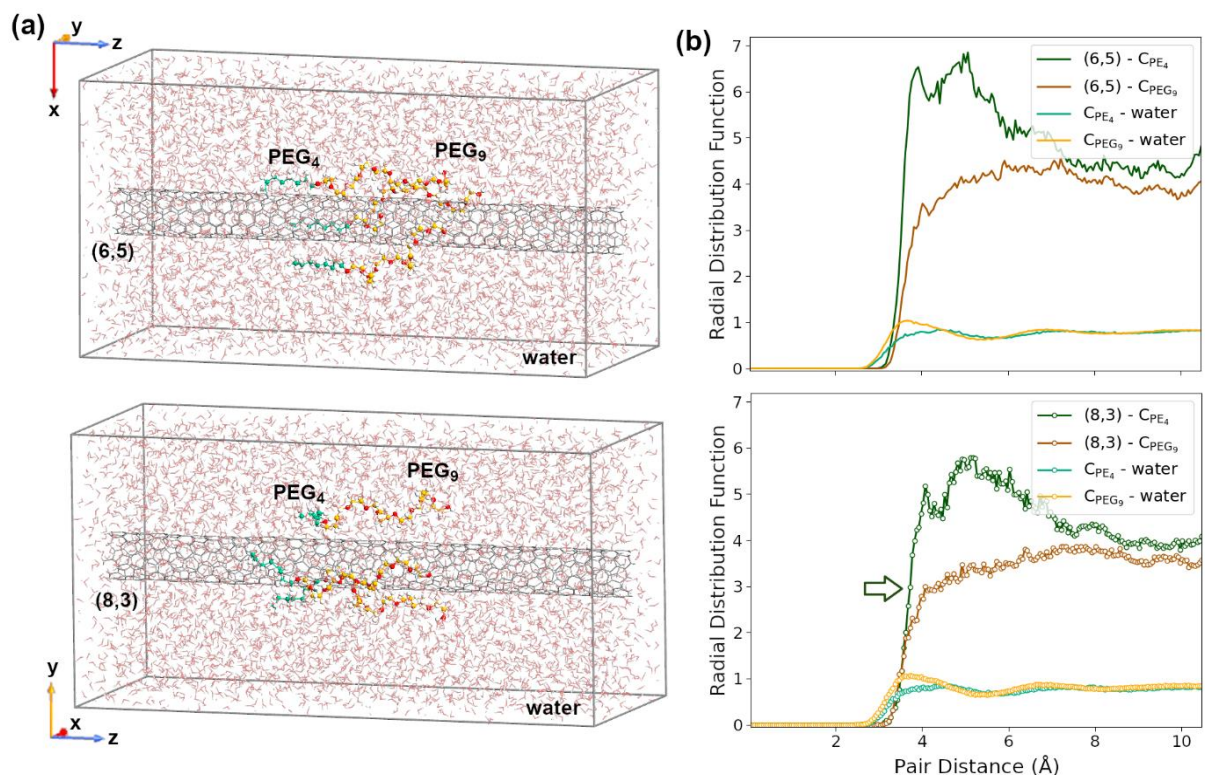

**Figure S5** MD/MC simulation of PE$_4$PEG$_9$–SWCNT interactions in aqueous solution. (a) Snapshots of the final configurations for simulation boxes containing short PEPEG polymers (comprising four PE and nine PEG units) interacting with two units of either a (6,5) SWCNT (top panel) or an (8,3) SWCNT (bottom panel), immersed in water (4491 or 4493 molecules, respectively). SWCNTs, solvent, and polymer molecules are represented using stick, line, and ball-and-stick models, respectively. For clarity, solvent molecules are rendered with increased transparency. Carbon atoms in the PE and PEG segments are shown in teal and yellow, respectively. Oxygen atoms are red, and hydrogen atoms are white. (b) Radial distribution functions (RDFs) for: SWCNT carbon–PE carbon, SWCNT carbon–PEG carbon, PE carbon–water, and PEG carbon–water. The trends observed for the shorter PEPEG polymer are consistent with those of the longer chain: the PE segment preferentially aligns closer to the SWCNT surface, while the PEG segment tends to remain further away, favoring solvation in water. Additionally, the first RDF peak corresponding to SWCNT–PE interaction is slightly shifted to larger distances for the (8,3) SWCNT compared to the (6,5) SWCNT. This shift, indicating a subtly weaker interaction between PE and the (8,3) chirality, is marked by a green arrow in the bottom panel.